\documentstyle{article}

\begin{document}

\title{Correlated hopping in the 1D Falicov-Kimball model}

\author{Z. Gajek and R. Lema\'{n}ski }
\date{\em Institute of Low Temperatures and Structure Research,\\
Polish Academy of Sciences, P.O.Box 1410, 50-950 Wroc\l aw, Poland }
\maketitle

\begin{abstract}
Both canonical and grand canonical phase diagrams of the one dimensional
Falicov-Kimball model (FKM) with correlated hopping are presented
for several values of the model parameters. Using the method of restricted
phase diagrams the system has been studied in the thermodynamic limit.
As compare to the conventional FKM, the diagrams exhibit a loss of the
particle-hole symmetry.
\vskip 0.3cm

\noindent PACS D5.30.Fk, 71.30+h
\end{abstract}

As the simplest, still non-trivial model of highly
correlated electron systems, the Falicov-Kimball model (FKM)
attracts growing attention among solid state physicists.
It can describe variety of the most intriguing cooperative phenomena,
as metal-insulator transition, mixed-valence phenomenon etc.
(see e.g. the review \cite{GM}).

The model deals with itinerant particles (electrons) that can hope
between nearest-neighbor sites. Some of
lattice sites are occupied by non-movable particles, playing a
role of ions or localized electrons; we call them "ions". The
only interaction in the system is the on-site, Coulomb-type
interaction between the electrons and the ions. The interaction
generates long-range correlations between the ions.

The model has been investigated thoroughly in nineties. Numerous
approximate results supplemented with some exact and rigorous statements
reported up to now provide a good basis for further extensions of the model
towards more realistic physical situations. These include
the discussed here model with correlated hopping, according to which
the electron hopping rate depends on occupations of relevant sites.

The Hamiltonian of the FKM with correlated hopping reads:
\begin{eqnarray}
H & =  - t\sum_{x}(c^\dagger _{x}c_{x+1}+c^\dagger _{x+1}c_{x})
\{1 - \alpha [w(x)+ w(x+1)-\gamma w(x) w(x+1)]\} \nonumber \\
& - U \sum_{x}w(x)c^\dagger _{x}c_{x},&
\label{hamq}
\end{eqnarray}
where $w(x)$ denotes the ion occupation number at site x (it takes a value 0 or 1),
$c^\dagger _{x}, c_{x}$ are the operators that create and annihilate an electron
at site $x$, respectively. The electron-ion coupling constant $U$ is chosen to be
positive, what corresponds to the attraction between the ions and the electrons.
Note that with the parametrization given in (\ref{hamq})
the hopping amplitudes can take the three following values: $t_{00}=t$ if an electron
hops between two empty sites, $t_{01}=t(1-\alpha )$ if it hops between one site
occupied by an ion and the other empty (obviously $t_{01}=t_{10}$), and finally
$t_{11}=t[1-\alpha (2-\gamma )]$ if it hops between two sites occupied by ions.
For $\alpha =0$ the Hamiltonian (\ref{hamq}) reduces to the conventional FKM
without correlated hopping.

The meaning of the correlated hopping parameters $\alpha$ and $\gamma$ depends on
a particular physical situation to be modeled. For instance, the $\alpha$ parameter
may originate from {\em bond-charge} repulsion, the mechanism originally discussed
in the frames of the extended Hubbard model \cite{Hubbard,Hirsch}.
Within the same microscopic picture the parameter $\gamma$ depends strongly on the
effective nuclear charge $Z$ and apparently decreases for larger $Z$ \cite{Hirsch}.

In general, recognition and understanding the mechanisms leading to the correlated
hopping, as well as its consequences are far from being satisfactory. In particular,
this concerns the problem of formation of stable phases. The present work turns towards
this direction for the simplest, one-dimensional case.

Various approaches known for the ordinary FK model have been adopted to its
extended version \cite{O}. Here we used the method of {\em restricted phase diagrams},
where infinite systems of periodic phases, whose period does not exceed some $r_{max}$
(here we took $r_{max}=7$), as well as their mixtures were considered. The Gibbs
potentials of all these periodic phases were calculated exactly \cite{RL},
so we were able to get the ground state phase diagrams (in a ($\mu_e$ , $\mu_i$) plane)
with a high precision. Then we mapped them onto the $(\rho_e, \rho_i)$ plane,
thus obtaining canonical phase diagrams. Details of the method and the calculation
procedures were published previously (see \cite{JLL,GJL}).

Our results are presented in Figs. 1 and 2, where we took the intermediate value of $U$
being equal to 1.6, $\alpha = 0$, 0.1, 0.2 and $\gamma = 0.0$, 0.5.
In Fig. 1 the restricted grand canonical phase diagrams are displayed,
whereas in Fig. 2 the corresponding canonical phase diagrams are shown.
Figs. 1(a) and 2(a) correspond to the simplest FKM, with no correlated hopping terms
($t_{00}=t_{01}=t_{11}=t$).
In this case the diagram is symmetric with respect to exchange between
sites occupied by the ions and those unoccupied (it has {\em the particle-hole
symmetry}). An extended analysis of that case was given previously \cite{GJL};
here we included the diagram only as a reference one.

If one "turns on" the correlated hopping in such a way that
$\alpha \neq 0$ and $\gamma =0$ -- see Fig 1(b), (c) and 2(b), (c),
then with an increasing $\alpha$ the diagrams become
more and more asymmetric. For $\alpha =0.2$ ($t_{00}=t$, $t_{01}=0.8 t$
and $t_{11}=0.6 t$) all periodic phases laying on the left
from the $\rho _i+\rho _e=1$ line disappear and their place take mixtures
of periodic neutral phases with the "empty" one (with free electrons and no ions).
Instead the the so-called {\em three-molecular} periodic phases \cite {GJL}
develop on the right hand side from the $\rho _i+\rho _e=1$ line.

On the other hand, if we increase the electron hopping amplitude between
two occupied sites $t_{11}$ from $0.6t$ to $0.7t$ (what corresponds
$\alpha =0.2$ and $\gamma =0.5$ -- see Figs. 1(d) and 2(d) ),
then those {\em three-molecular} phases are suppressed.

A very brief analysis of the displayed phase diagrams show
their considerable sensitivity to a variation
of the correlated hopping parameters. In particular, the regions
enclosed within the triangles close to the lower-left and upper-right corners
of the canonical diagrams, where {\em the segregated phase}
(a mixture of an empty lattice with free electrons
and the fully occupied lattice with a number of electrons) is stable,
clearly depends on values of the correlated hopping amplitudes (see Fig. 2).
The above preliminary results confirm conjectures already published several
years ago (e.g. \cite{Hirsch}), that the correlated hopping plays
an important role and should be taken into
account if one intends describe propely physical properties of the systems.

\vspace{0.2cm}
\noindent{\bf Acknowledgements.} We thank Janusz J\c{e}drzejewski for valuable
and fruitful discussions. Support from the Polish Research Committee (KBN)
under the Grant No. 2 P03B 131 19 is greatly acknowledged.

\newpage
\begin{center}
Figure captions
\end{center}

\noindent Fig. 1. The restricted grand canonial phase diagrams for $U = 1.6 t$
and the following four sets of correlated hopping parameters:
(a) $\alpha =0$ and $\gamma =0$; (b) $\alpha =0.1$ and $\gamma =0$;
(c) $\alpha =0.2$ and $\gamma =0$; (d) $\alpha =0.2$ and $\gamma =0.5$.
Areas displayed on the diagrams are shadowed according to ion densities
of representing them phases. Each line in an upper part of the segment
(a), (b), (c) or (d) of Figure 1 contains three characteristics of those
phases, that appear in the diagram: the ratios in the square brackets stand for
their ion densities (while the denominators are equal to the periods of corresponding
phases); the sequences of circles and dots (corresponding to the ions and
empty sites, respectively) represent their unit cells; shadowed stripes
show energy intervals where density of states of the phases are positive.
\vskip 0.3cm

\noindent Fig. 2. The restricted canonical phase diagrams for $U = 1.6 t$
and for the same sets of correlated hopping parameters as those given in Fig. 1.
The black spots represent the periodic phases (whose period is at least 2).
The straight line segments join those spots whose corresponding phases
touch each other on the ground canonical phase diagram.
The points located on a segment represent the mixtures of the two
periodic phases that correspond to the ends of the segment.
The black spots located on the lines $\rho_{i}=0$ or $\rho_{i}=1$
show the minimal and maximal electron densities of the full phases
that form mixtures with the same periodic phase.

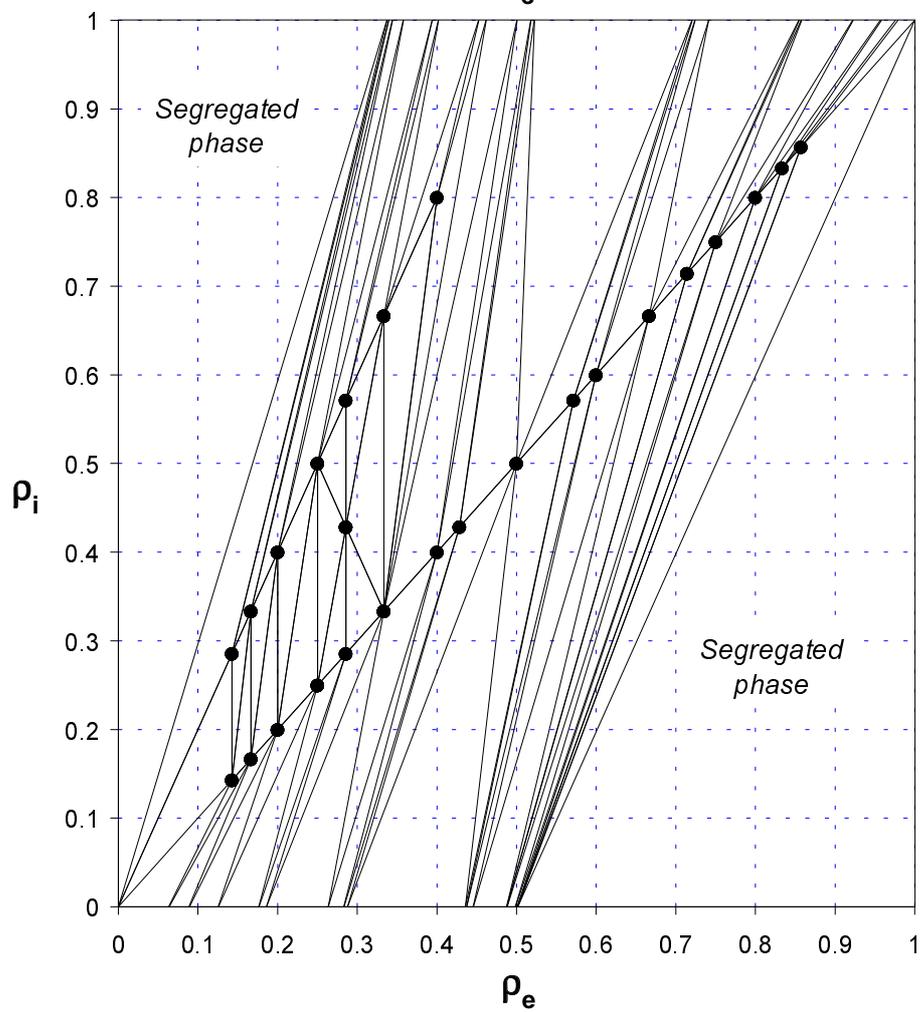
\end{document}